# Generalization of the circular dichroism from metallic arrays that support Bloch-like surface plasmon polaritons

X. Guo[a)], C. Liu, and H.C. Ong[b)]

Department of Physics, The Chinese University of Hong Kong, Shatin, Hong Kong, People's Republic of China

The broken mirror symmetry in subwavelength photonic systems has manifested many interesting chiroptical effects such as optical rotation and circular dichroism. When such systems are placed periodically in a lattice form, in addition to intrinsic chirality, extrinsic chirality also takes part, and the overall effect depends not only on the basis and lattice but also the excitation configuration. Here, we study planar chiral nanohole arrays in square lattice that support Bloch-like surface plasmon polaritons (SPPs) and clarify how the system geometry and the excitation contribute to circular dichroism. By using temporal coupled mode theory (CMT), the dissymmetry factor and the scattering matrix of the arrays are analytically formulated. Remarkably, we find the dissymmetry factor depends only on the coupling polarization angle and the in-coupling phase difference between the $p$- and $s$-polarizations. Besides, the upper limit of the dissymmetry factor at $\pm 2$ can be reached simply by orienting the lattice of the arrays for properly exciting the Bloch-like SPPs and at the same time making the basis mimic two orthogonal and relatively displaced dipoles, demonstrating the interplay between extrinsic and intrinsic chirality. The models have been verified by numerical simulations and experiments, yielding the dissymmetry factors to be 1.82 and 1.55, respectively, from the proposed dual slot system.

[a)] Present address: Physics Department, University of Michigan, Ann Arbor, MI 48108, USA.
[b)] hcong@phy.cuhk.edu.hk

# I. INTRODUCTION

In chemistry, a molecule is called chiral when it is not superimposable with its mirror image [1]. One of the most interesting properties of chiral molecules is their interaction with light. For example, they exhibit different absorptions under the illumination of right (R) and left (L) circularly polarized (CP) lights, exhibiting the well-known circular dichroism (CD) [1]. Because the CD arising from chiral molecules and their mirror images are opposite in sign, it becomes a very popular technique in differentiating the handedness of the enantiomers [2]. Recently, such idea has been employed in photonics where the geometry of subwavelength optical systems is designed to break the mirror symmetry [3-9]. Remarkably, the light-matter interactions arising from chiral photonic systems follow their molecular counterparts very well. Both optical rotation (OR), which tilts the linear polarization of light about its optical axis, and circular dichroism are observed in chiral photonic systems [10-18]. As a result, there have been intensive efforts devoted to designing and implementing nanoscale chiral systems for optimizing the chiroptical effects [3-4,19-24].

However, rationally designing the chiral systems is not a trivial task. In fact, most of the studies on chiral photonics involve complex system geometry that cannot be analytically simplified. This is because those systems are no longer considered as point dipoles and expansion to higher orders are necessary [25-27]. Therefore, the size and shape of the system play a vital role in governing the light-matter interactions [14,26,28]. The situation becomes even more complicated when such systems are placed in a lattice form where, in addition to the basis, both the lattice and the excitation configuration should also be considered properly [29-32]. Currently, numerical electrodynamic simulations are usually performed to identify the core structures of the system and then fine tune them in a step by step manner [14,26,33-37]. Likewise, experiments mostly rely on a trial-and-error approach and are supplemented with simulations when necessary [13,38-41]. However, these two methods are very time consuming and sometimes work as a black box, which does not reveal much of the physics behind. It is always desirable if the physical mechanisms of chiral photonics can be generalized despite the diversity of system geometry and excitation.

In general, when resonances are involved, both CD and OR are the consequences of the in/out-couplings of lights to/from the resonator. How the resonance is excited and dissipated under different polarizations determine the light absorption and scattering as well as the phase difference between the outgoing polarizations. For example, an ideal CD requires the resonator to be completely absorbing for one circular polarization but scattering for the other [3-4,24]. On the other hand, in analogy to a half wave plate, light polarization gets flipped when one of the linear polarizations encounters a π phase shift upon the excitation of the resonator [42]. Therefore, if one can generalize the complex in- and out-coupling constants under different system geometries and excitation conditions, the chiroptical effects of some specific systems may be controlled at will.

Here, we attempt to accomplish such task on planar chiral nanohole arrays where Bloch-like surface plasmon polaritons (SPPs) are supported. In particular, we combine temporal coupled mode theory (CMT), electrodynamic simulation, and polarization- and angle-resolved reflectivity spectroscopy together to formulate the dissymmetry factor and the scattering matrix of 2D L-shape Au nanohole arrays. We find the complex in-coupling constant, which consists of the coupling polarization angle and phase shift, play a significant role in determining the resulting CD and they are found to be strongly dependent on the system geometry and excitation configuration. The dissymmetry factor can be maximized to ±2 when the in-coupling polarization angle is ±45° and the difference between the $p$- and $s$-phase shifts is ±90°. More importantly, such conditions can be accomplished by decoupling the L-shape basis into two relatively displaced orthogonal slots and at the same time orientating the incident plane and the lattice properly to excite the SPPs. The optimization manifests the interplay between extrinsic and intrinsic chirality, which are carefully controlled through the rational design of the coupling constants.

## II. EXPERIMENT

We have fabricated 2D L-shaped nanohole arrays by focused ion beam (FIB). First, 300 nm thick Au films are deposited on glass substrates by radio frequency magnetron sputtering. Then, different array patterns with area of approximately 0.01 mm$^2$ are milled on the Au films by FIB. The plane-view scanning electron microscopy (SEM) image of one of the samples is shown in the inset of Fig. 1(a), showing it has period $P$ = 550 nm, a long and a short arm with width = 125 nm and lengths = 400 and 250 nm. Since the Au film is optically thick, the sample has no transmission. Once the sample is ready, it is transferred to a homebuilt optical microscope where angle- and polarization-resolved reflectivity measurement can be performed [43]. The setup is shown in Fig. 1(a). Briefly, a broadband supercontinuum laser is collimated and then passed through a set of polarizers, wave plates, and lenses before being focused onto the back focal plane (BFP) of a 100X objective lens with numerical aperture = 0.9. By displacing the focused spot across the BFP of the objective lens using a motorized translation stage, the light exiting from the objective lens is collimated again and the incident polar angle $\theta$ onto the sample is given by $\sin\theta = d/f$, where $d$ is the distance between the focused spot and the optical axis of the objective lens and $f$ is the focal length of the objective [44]. In addition, the azimuth angle $\phi$ can be varied by a motorized rotation sample stage. The light reflected from the sample will then be collected by the same objective lens and passes through another set of analyzers and lenses before being detected by a spectrometer-based CCD detector. For CD measurement, we focus on the modes that support only specular reflection so that the dissymmetry factor $g$ is given as [1]:

$$g = 2\left(\frac{A_{RCP} - A_{LCP}}{A_{RCP} + A_{LCP}}\right) = 2\left(\frac{R_{LCP} - R_{RCP}}{2 - R_{RCP} - R_{LCP}}\right), \quad (1)$$

where $A_{LCP/RCP}$ and $R_{LCP/RCP}$ are the absorptions and reflectivities taken under LCP and RCP lights.

## III. RESULTS

The polarization- and angle-resolved reflectivity mappings of the L-shape nanohole array are presented here. The measurement configuration is shown in Fig. 1(b). At $\theta = 45°$, the $\phi$-resolved reflectivity mappings taken under different polarizations, $R_{pp}$, $R_{ps}$, $R_{sp}$, and $R_{ss}$, are illustrated in Fig. 2(a)-(d). The first and second subscripts in $R$ indicate the collection and incident polarizations. We see from the $R_{pp}$ and $R_{ss}$ mappings that multiple low reflection dispersive bands are observed, and they can be identified by the SPP phase-matching equation given as [32]:

$$\vec{k}_{SPP} = \left(\frac{2\pi}{\lambda}\sin\theta\cos\varphi + \frac{n_x 2\pi}{P}\right)\hat{x} + \left(\frac{2\pi}{\lambda}\sin\theta\sin\varphi + \frac{n_y 2\pi}{P}\right)\hat{y}, \quad (2)$$

where $\vec{k}_{SPP}$ is the propagation vector of the SPP mode defined with respect to the $\Gamma$-X direction and $(n_x, n_y)$ are the indices specifying the Bragg scattering order. The propagation constant of SPPs $|\vec{k}_{SPP}| = \frac{2\pi}{\lambda}\sqrt{\frac{\varepsilon_{Au}}{1+\varepsilon_{Au}}}$, where $\varepsilon_{Au}$ is the dielectric constant of gold [45]. Therefore, the reflection dips are the lowest order (-1,0), (0,-1) and (1,0) SPPs in addition to the (-1,-1) and (1,-1) higher order modes at shorter wavelength. We find that the mappings are not symmetric with respect to $\phi = 90°$ due to the presence of the chiral basis that breaks mirror symmetry. In particular, the reflectivity profiles from 35° to 90° and those from 90° to 145° are dramatically different. For example, at the 45° and 135° cross points where two propagating SPPs interact together to form a pair of bright and dark modes as well as a plasmonic band gap, one can see the gap is negligibly small at 45° but noticeable at 135°, indicating different coupling strengths [46]. We also see strong polarization conversion in the $R_{ps}$ and $R_{sp}$ mappings although their degrees of conversion are different, demonstrating the strong anisotropy introduced by the chiral basis. It is noted that both (-1,0) and (1,0) SPP modes exhibit stronger polarization conversion than the (0,-1) mode.

We then measure the absorption mappings of the array taken under LCP and RCP lights in Fig. 3(a) & (b) for (-1,0), (0,-1) and (1,0) SPPs. These modes are chosen because they support only the specular reflection so that the dissymmetry factor $g$ can be determined easily by Eq. (1). Again, the mappings are not symmetric with respect to $\phi = 90°$. More importantly, the absorptions from two CP lights are different, indicating the presence of CD. We then calculate $g$ along the lowest SPP bands as a function of $\phi$ in Fig. 3(c). We observe $g$ is almost zero when $\phi$ is close to 80° but exhibits a monotonic behavior before reaching the highest negative and positive $g$ at the 45° and 135° cross points. In the following, we will focus our effort in explaining the behavior of $g$ for the nondegenerate (0,-1) SPPs excluding the cross points.

## IV. FORMULATION OF DISSYMMETRY FACTOR

We attempt to formulate the $g$ of the chiral plasmonic system within the framework of CMT [32,47-49]. We consider the system that supports one single resonance and two input-output ports for $p$- and $s$-polarizations. It therefore mimics an optically thick system that supports the lowest (-1,0) SPPs where only specular reflection is present. The dynamics of the mode amplitude $a$ can be written as [47]:

$$\frac{da}{dt} = -i\omega_o a - \frac{\Gamma_t}{2}a + \kappa_p s_{+p} + \kappa_s s_{+s}, \qquad (3)$$

where $\omega_o$ is the resonant angular frequency, $\Gamma_t$ is the total decay rate, which is the summation of the absorption and radiative decay rates, i.e. $\Gamma_{abs} + \Gamma_{rad}$, $\kappa_{p/s}$ are the complex in-coupling constants for $p$- and $s$-polarizations, and $s_{+p/s}$ are the $p$- and $s$-polarized incident power amplitudes. It is noted that $\kappa_p = \sqrt{\Gamma_{rad}} \cos\alpha\, e^{i\delta_p}$ and $\kappa_s = \sqrt{\Gamma_{rad}} \sin\alpha\, e^{i\delta_s}$, where $\alpha$ is the coupling polarization angle and $\delta_p$ and $\delta_s$ are the in-coupling phase shifts [32,50]. Physically, $\alpha$ is the polarization angle, defined with respect to the $p$-polarization, where the excitation of SPPs is optimal [32,50]. In fact, Eq. (3) has been applied to 2D square lattice circular hole arrays in which $\alpha$ is related to the propagation direction of SPPs provided by Eq. (2). In addition, $\delta_p$ and $\delta_s$ are always equal due to the circular nanoholes [32].

Modification of Eq. (3) is necessary when extended to chiral basis. For broken mirror symmetry, in analogy to planar birefringence, we expect both $\alpha$ and $\delta_{p/s}$ to be amended. First, empirically, $\delta_p$ and $\delta_s$ are no longer identical, leading to different coupling phase shifts under the *p*- and *s*-polarizations [51-53]. Second, in addition to the lattice contribution, the anisotropy of the basis introduces additional dipole contribution to $\alpha$, giving rise to $\alpha = \alpha_{lattice} + \alpha_{basis}$, which will be discussed later. When on resonance where $\omega = \omega_0$, by solving Eq. (3), we have $|a|^2 = \frac{2\Gamma_{rad}}{\Gamma_t^2}\left(1 + \sin 2\alpha \cos(\delta_p - \delta_s - \gamma)\right)$, where $\gamma$ is the phase difference between the *p*- and *s*-incident lights.

Knowing from CMT that the absorption $A = \Gamma_{abs}|a|^2/(|s_{+p}|^2 + |s_{+s}|^2)$ and the LCP and RCP lights with $\gamma = \pm\pi/2$ are $\sqrt{\frac{1}{2}}\begin{bmatrix}1\\i\end{bmatrix}$ and $\sqrt{\frac{1}{2}}\begin{bmatrix}1\\-i\end{bmatrix}$, $A_{RCP}$ and $A_{LCP}$ can be expressed as $\frac{2\Gamma_{rad}\Gamma_{abs}}{\Gamma_t^2}\left(1 \mp \sin 2\alpha \sin(\delta_p - \delta_s)\right)$. Then, we reach our main result as:

$$g = -2\sin 2\alpha \sin(\delta_p - \delta_s), \tag{4}$$

which arises primarily from the interference between two *p*- and *s*-excited SPPs. From Eq. (4), we see *g* varies between ±2 and depends solely on $\alpha$ and $\delta_p - \delta_s$. In particular, if $\alpha = \pm 45°$ and $\delta_p - \delta_s = \pm\pi/2$, *g* is the highest. We explain this by examining the interference term $\mp\sin 2\alpha \sin(\delta_p - \delta_s)$ which depends on the amplitudes and the phase difference between the *p*- and *s*-excited SPP waves. For a given CP, when $\alpha = \pm 45°$, both *p*- and *s*-waves are excited equally. At the same time, if $\delta_p - \delta_s = \pm\pi/2$, we see one CP will give constructive interference with $A = 4\Gamma_{rad}\Gamma_{abs}/\Gamma_t^2$ whereas the other yields destructive interference with $A = 0$. In other words, the system is absorbing for one CP but completely reflective for the other. Eq. (4) also explains why circular nanohole arrays do not give rise to any CD [32]. For circular basis, despite $\alpha \neq 0°$ under certain excitation configurations, $\delta_p - \delta_s$ is always equal to zero due to the lack of anisotropy. Finally, as both $\alpha$ and $\delta_p - \delta_s$ are system dependent, one can follow Eq. (4) to rationally design *g*.

It is noted that the picture described here is consistent with the recent work by Tang and Cohen [54-55]. In their work, the absorption taken under CP excitations is divided into two parts as $\frac{\omega}{2}\left(\alpha''|\vec{E}|^2 + \chi''|\vec{B}|^2\right)$ and $G''\omega Im(\vec{E}^*\cdot\vec{B})$, where $\vec{E}$ and $\vec{B}$ are the electric and magnetic fields and $\alpha''$, $\chi''$, and $G''$ are the imaginary part of the electric polarizability, the magnetic susceptibility, and the isotropic mixed electric-magnetic dipole polarizability. The first part is related to the electric and magnetic energy densities whereas the second is associated with optical chirality. CD thus arises from the difference between the energy density and the optical chirality from two CPs. For our case, although the Bloch-like SPP wave carries spin angular momentum (SAM), it only has transverse SAM that does not contribute to any optical chirality [56,57]. Therefore, the CD from our arrays is solely due to the difference in energy density, i.e. $|a|^2$, which agrees well with our CMT.

## V. FORMULATION OF SCATTERING MATRIX

We then formulate the scattering matrix $S$ of the system. We note that the L-shape basis breaks not only the left-right mirror symmetry but also that of forward-backward under $\vec{k}$ and $-\vec{k}$ incident light directions. The scattering matrix $S$, which is defined as $\begin{bmatrix} s_{-p} \\ s_{-s} \end{bmatrix} = S \begin{bmatrix} s_{+p} \\ s_{+s} \end{bmatrix}$, where $s_{-p/s}$ are the $p$- and $s$-polarized outgoing power amplitudes, is no longer symmetric but fulfills the following condition [58]:

$$S(\omega, \vec{k}) = S^T(\omega, -\vec{k}), \quad (5)$$

where the superscript $T$ is the transpose of the matrix. Under forward $\vec{k}$ incidence, $s_{-p/s}$ can be expressed as:

$$\begin{bmatrix} s_{-p}^k \\ s_{-s}^k \end{bmatrix} = C \begin{bmatrix} s_{+p}^k \\ s_{+s}^k \end{bmatrix} + a^k \begin{bmatrix} d_p^k \\ d_s^k \end{bmatrix}, \quad (6)$$

where $C = \begin{bmatrix} \tilde{r}_{pp}^k & \tilde{r}_{ps}^k \\ \tilde{r}_{sp}^k & \tilde{r}_{ss}^k \end{bmatrix}$ is the complex direct reflection matrix and $d_{p/s}$ are the complex out-coupling constants. For $C$, the first and second subscripts again denote the collection and incident polarizations. Substitute $a^k$ from Eq. (3) to Eq. (6), we have:

$$\begin{bmatrix} s_{-p}^k \\ s_{-s}^k \end{bmatrix} = \left[ C + \frac{1}{i(\omega_o - \omega) + \Gamma_t/2} \begin{pmatrix} d_p^k \kappa_p^k & d_p^k \kappa_s^k \\ d_s^k \kappa_p^k & d_s^k \kappa_s^k \end{pmatrix} \right] \begin{bmatrix} s_{+p}^k \\ s_{+s}^k \end{bmatrix} = S(\omega, \vec{k}) \begin{bmatrix} s_{+p}^k \\ s_{+s}^k \end{bmatrix}, \quad (7)$$

which gives the scattering matrix $S(\omega, \vec{k})$. As is given in the Appendix, when combining Eq. (5) and (7) together with time reversal symmetry and conservation of energy [47-49], we finally have:

$$S(\omega,\vec{k}) = \begin{bmatrix} r_{pp}^k & r_{ps}^k \\ r_{sp}^k & r_{ss}^k \end{bmatrix}$$

$$= \begin{bmatrix} \tilde{r}_{pp}^k + \dfrac{\Gamma_{rad} \cos\alpha_k^{in} \cos\alpha_k^{out} e^{i(\delta_p^{k,in} + \delta_p^{k,out})}}{i(\omega_o - \omega) + \Gamma_t/2} & \tilde{r}_{ps}^k + \dfrac{\Gamma_{rad} \sin\alpha_k^{in} \cos\alpha_k^{out} e^{i(\delta_s^{k,in} + \delta_p^{k,out})}}{i(\omega_o - \omega) + \Gamma_t/2} \\ \tilde{r}_{sp}^k + \dfrac{\Gamma_{rad} \cos\alpha_k^{in} \sin\alpha_k^{out} e^{i(\delta_p^{k,in} + \delta_s^{k,out})}}{i(\omega_o - \omega) + \Gamma_t/2} & \tilde{r}_{ss}^k + \dfrac{\Gamma_{rad} \sin\alpha_k^{in} \sin\alpha_k^{out} e^{i(\delta_s^{k,in} + \delta_s^{k,out})}}{i(\omega_o - \omega) + \Gamma_t/2} \end{bmatrix}, \quad (8)$$

where the superscripts *in* and *out* define the in- and out-couplings, or the incident and reflection sides. We see, when under $\vec{k}$ direction, the reflection coefficients $r$ in the matrix result from the interference between direct reflection and the radiation damping from SPPs. It is also noted the incident and reflection polarization angles $\alpha$ and phase shifts $\delta_{p/s}$ are different. As $-\vec{k}$ is $\vec{k}$ swiveled by 180°, for simplicity, we omit $k$ in the expressions if not necessary.

## VI. NUMERICAL SIMULATIONS

We validate the CMT models by electrodynamic simulations. We have performed finite element method (FEM) by COMSOL and finite-difference time-domain (FDTD) by Lumerical to simulate the complex reflection coefficient $r$, i.e. amplitude and phase, spectra of the (0,-1) SPPs from a L-shape Au array. The unit cell is shown in the inset of Fig. 4(a) and has period $P = 550$ nm, hole depth $H = 60$ nm, long and short arms $a = 400$ nm and $b = 250$ nm as well as arm width $w = 125$ nm. Bloch boundary condition is used on four sides. At a fixed polar incident angle $\theta = 45°$, we calculate by FEM the spectra as a function of $\phi$ under different incident and collection polarizations and two of them taken at $\phi = 120°$ and $300°$, i.e. forward and backward incidences, are shown in Fig. 4(a)-(h) and others are given in the Supplementary Information [59]. Comparing the complex $r_{ps}$ and $r_{sp}$ taken under two opposite incidences from Fig. 4(b), (c) and (f), (g), we see $r_{ps}^{k(-k)} \neq r_{sp}^{k(-k)}$, demonstrating $S(\omega,\vec{k})$ is not symmetric. In addition, it is also clear that $r_{pp}^k = r_{pp}^{-k}$, $r_{ss}^k = r_{ss}^{-k}$, and $r_{ps}^{k(-k)} = r_{sp}^{-k(k)}$, in agreement with Eq. (5) and (8).

The amplitude and phase profiles are then best fitted by Eq. (8) to determine $\alpha^{in}$ and $\delta_p^{in} - \delta_s^{in}$ which are plotted in Fig. 5(a) and (b) against $\phi$. Some examples are shown in Fig. 4 as solid lines, indicating reasonably good fits. Alternatively, we assume the incident wave has a functional form of $\begin{bmatrix} \cos\alpha \\ \sin\alpha e^{i\delta} \end{bmatrix}$ such that $\alpha$ and $\delta$ define the incident polarization angle and phase. We then systematically vary $\alpha$ and $\delta$ while at the same time monitoring the SPP near-field intensity. The $\alpha$ and $\delta$ that produce the strongest field intensity are also plotted in Fig. 5(a) & (b) and they agree with $\alpha^{in}$ and $\delta_p^{in} - \delta_s^{in}$ deduced from CMT. We thus interpret $\alpha^{in}$ and $\delta_p^{in} - \delta_s^{in}$ as the best incident polarization angle and phase difference that maximize the energy transfer from far-field to SPP near-field [32]. We note $\delta_p^{in} - \delta_s^{in}$ encounters discontinuity near $\phi = 80°$. However, given $\alpha^{in} \approx 0°$ where the best polarization is *p*-polarized and no *s*-polarized light can couple to the resonance, the in-coupling phase difference thus is ill-defined.

We also calculate $g$ as a function of $\phi$ by using Eq. (4) and $\alpha^{in}$ and $\delta_p^{in} - \delta_s^{in}$ obtained from the field manipulation method, the best fits as well as the direct FEM simulations taken under LCP and RCP lights. Their results are plotted in Fig. 5(c), showing good agreement with discrepancy in less than 10%. The experimental $g$ from Fig. 3(c) is overlaid in Fig. 5(c), showing it is consistent with the simulations that $g$ is close to zero at $\phi \sim 80°$ and exhibits similar monotonic behavior. Discrepancies are observed between experiment and simulation, particularly at small and large $\phi$. However, discrepancies are always present due to the imperfection of the sample preparation, surface roughness, sample nonuniformity, etc, that are not easily removed completely. We also best fit the spectra extracted from Fig. 2 to determine $\alpha^{in}$ of the (0,-1) SPPs as a function of $\phi$ in Fig. 5(a) and they are consistent with the simulation results. We therefore conclude both Eq. (4) & (8) describe our system properly.

## VII. DISCUSSION ON α

Before we move to rationally design the system to optimize $g$, it is essential to understand the physics behind $\alpha^{in}$. For circular nanoholes, $\alpha^{in}$ is found to depend solely on the incident angle $\theta$ and the SPP propagation direction $\rho$ and is given as [32]:

$$tan\,\alpha^{in} = cos\,\theta\,tan\,\rho, \tag{9}$$

where $\rho$ is defined with respect to the incident plane based on Eq. (2). However, when anisotropy is introduced, $\alpha^{in}$ requires modification. We tackle this by comparing the $\alpha^{in}$ deduced from Eq. (9) and the calculations from Fig. 5(a) and observe a large discrepancy. The difference is taken and plotted in Fig. 5(d), showing the discrepancy is almost constant at 9-10° in the middle in which the (0,-1) SPP mode is non-degenerate but begins to increase and decrease to 25° and 1° when approaching to the cross-points at 45° and 135°. Such constant discrepancy implies there exist an additional factor, which arises from the basis, that takes part in yielding the overall $\alpha^{in}$. The same difference plot from another L-shaped nanohole array with the same basis but $P$ = 800 nm taken at $\theta$ = 5° is provided in the Supplementary Information, supporting our observation [59]. Based on this, we propose $\alpha^{in}$ is contributed from both the lattice and basis as:

$$\alpha^{in} = \alpha^{in}_{lattice} + \alpha^{in}_{basis}. \tag{10}$$

Specifically, while $\alpha^{in}_{lattice}$ arises solely from the lattice following Eq. (9), $\alpha^{in}_{basis}$ strongly depends on the size, shape and orientation of the basis. In particular, since Eq. (9) is formulated under the assumption that the best excitation occurs when the incident polarization aligns with the longitudinal electric field of SPPs, any deviation of $\alpha$ from $\alpha^{in}_{lattice}$ is most likely coming from the emergence of a field component that is orthogonal to the SPP propagation [32]. Unlike the circular basis, the anisotropy introduced by the chiral L-shaped bases deviates the direction of the dipole moment and gives rise to an in-plane transverse field that effectively perturbs $\alpha$.

To verify the idea, we have simulated two series of L-shape arrays with different arm lengths and orientations. For the first series, four arrays with $P$ = 800 nm and arm lengths being 0, 100, 300, and 500 nm as shown in Fig. 6(a) are simulated at $\theta$ = 5° along the Γ-X direction. Their $\alpha^{in}$ are plotted in Fig. 6(b). Provided $\rho$ = 0°, or $\alpha^{in}_{lattice}$ = 0°, we see when the basis is rectangular, i.e. arm length = 0 nm, in which the dipole moment is pointing along the Γ-X direction, it does not give rise to any $\alpha^{in}_{basis}$. However, as the arm length becomes longer and begins to swivel the dipole moment away from the Γ-X direction, $\alpha^{in}_{basis}$ increases correspondingly. We also calculate the electric field component perpendicular to the propagation direction of SPPs, i.e. the Γ-X direction, and find the strength increases consistently with increasing the arm length, as shown in Fig. 6(c). The second series further confirms the relationship between $\alpha^{in}_{basis}$ and the direction of the dipole moment. For the second series, we fix the geometry of the L-shape but vary its orientation. The basis is gradually rotated clockwise with respect to the Γ-X direction in Fig. 7(a) and the corresponding $\alpha^{in}$ are determined in Fig. 7(b). We find $\alpha^{in}$ changes monotonically as the L-shape basis rotates, indicating the basis rotation simply swivels a fixed dipole moment gradually away from the Γ-X direction, thus increasing $\alpha_{basis}^{in}$. Again, the corresponding orthogonal electric field strength is plotted in Fig. 7(c) and it shows an identical pattern as that of $\alpha^{in}$. As a result, we conclude an additional $\alpha_{basis}^{in}$ evolves when the basis becomes anisotropic. More importantly, we also see from Eq. (10) the subtlety of $\alpha^{in}$ arises from the interplay between intrinsic and extrinsic chirality. While $\alpha_{lattice}^{in}$ arises solely from the extrinsic chirality where only the relative orientation between the lattice and incident plane is relevant, $\alpha_{basis}^{in}$ is influenced by the strength of the dipole moment due to the symmetry breaking as well as its orientation with respect to the incident plane.

## VIII. RATIONAL DESIGN OF DUAL SLOT SYSTEM

We are now in the position to rationally design the plasmonic arrays to achieve the best $g$ by optimizing $\alpha^{in}$ and $\delta_p^{in} - \delta_s^{in}$ close to ±45° and ±π/2 following Eq. (4). We realize that when properly placed, the L-shape nanohole can actually be pictured as two orthogonal slots with each excited independently by a linear polarization [60]. As a result, the nanohole can be decoupled into two separate slots, with each one exciting a dipole moment perpendicular to itself, as given in the Supplementary Information [59]. Therefore, the schematic of the basis is proposed in Fig. 8(a) with two perpendicular rectangular slots with the same slot width and depth = 150 and 100 nm and lengths $a$ and $b$ separated by $s$. The whole basis is oriented by $\psi$ to align with the incident plane.

The design process is divided into two steps for $\alpha^{in}$ and $\delta_p^{in} - \delta_s^{in}$. First, we understand $\alpha^{in}$ consists of $\alpha_{lattice}^{in}$ and $\alpha_{basis}^{in}$, which provide us a great degree of freedom to tuning $\alpha^{in}$ to ±45°. On one hand, $\alpha_{lattice}^{in}$ can be adjusted by following Eq. (9), orienting the lattice with respect to the incident plane. On the other hand, $\alpha_{basis}^{in}$ is governed by the lengths $a$ and $b$ as well as the orientation $\psi$. $\alpha^{in}$ is also expected to be weakly dependent on $s$ as the separation between two slots may affect the resulting dipole moment. As an illustration, we calculate $\alpha^{in}$ from a series of dual slot systems with $P$ = 800 nm, $b$ = 300 nm, $s$ = 100 nm and $\psi$ = 60° taken under $\theta$ = 10° and $\phi$ = -120° so that $\alpha_{lattice}^{in}$ is roughly close to 34°. The slots are positioned in parallel and perpendicular to the incident plane so that they can be independently excited by $p$- and $s$-polarizations. We then fine tune $a$ from 200 to 400 nm in Fig. 8(b). It is found that $\alpha^{in}$ increases consistently with increasing $a$, verifying that $\alpha_{basis}^{in}$ is controlled by the relative strengths of the two orthogonal dipole moments with a larger dipole moment pointing in the direction perpendicular to the incident plane at large $a$. We also calculate $\alpha^{in}$ as a function of $s = 25$ to $100$ nm from another series of dual slot systems in Fig. 8(c) with the same $P$, $\psi$, and $b$ but the length $a$ is fixed at 300 nm under the same excitation condition. A weak dependence of $\alpha^{in}$ on $s$ is observed. Noted from Fig. 8(b) & (c) that $\alpha^{in}$ is close to 45° when $a = 300$ nm and $s = 30$ nm.

Once the $\alpha^{in}$ is designed properly, we then move on to study the dependence of $\delta_p^{in} - \delta_s^{in}$ on the geometry. As $\delta_p^{in} - \delta_s^{in}$ is the phase difference arising from the $p$- and $s$-excited SPP waves, it can be controlled by spatially displacing two slots to introduce a dynamic phase difference between them. Therefore, varying $a$ and $s$ can be used for such introduction. We determine the $\delta_p^{in} - \delta_s^{in}$ from the previous two series of arrays in Fig. 8(b) & (c). The results show $\delta_p^{in} - \delta_s^{in}$ varies almost linearly with both $a$ and $s$. $\delta_p^{in} - \delta_s^{in}$ is close to 90° at $a = 300$ nm and $s = 30$ nm. Based on the $\alpha^{in}$ and $\delta_p^{in} - \delta_s^{in}$, we calculate the corresponding $g$ in Fig. 8(c) & (e) and find the highest $g$ from two series reach 1.5 and 1.82.

Following the simulation results, we have fabricated a dual slot array by FIB as a demonstration. The plane-view SEM image of the system is shown in Fig. 8(a) and it shows the array has $P = 800$ nm, slot width = 150 nm, slot lengths $a$ and $b = 300$ nm, and $s = 30$ nm. At $\theta = 10°$ and $\phi = -120°$, the absorption spectra are taken under RCP and LCP lights and are plotted in Fig. 8(f). One sees the absorption peak observed at $\lambda = 933$ nm indicating the (0,-1) SPP mode and the absorption taken under RCP is much stronger than that under the LCP counterpart. The $g$ is then deduced from the figure showing 1.55, which is slightly smaller than the expected 1.82. The discrepancy is very likely due to structural imperfection, especially the slot depth that is very difficult to be fabricated precisely by FIB. Finally, it is noted that the RCP absorption only reaches 0.21. To further boost the absorption, we see from $A_{RCP/LCP} = \dfrac{2\Gamma_{rad}\Gamma_{abs}}{\Gamma_t^2}\left(1 \mp \sin 2\alpha \sin\left(\delta_p - \delta_s\right)\right)$ that other than $\alpha$ and $\delta_p - \delta_s$, $A$ will reach unity in constructive interference when $\Gamma_{rad} = \Gamma_{abs}$, which is also known as critical coupling or complete absorption [61].

## IX. CONCLUSION

In conclusion, we have experimentally and theoretically studied CD from 2D L-shaped nanohole arrays that support Bloch-like SPPs. Angle-resolved CD spectroscopy has shown the CD has strong dependence on system geometry and orientation. CMT and numerical simulations have been used to analytically formulate the dissymmetry factor $g$ and the scattering matrix $S$ of the arrays. In particular, $g$ is found only to depend on two parameters, which are the best incident polarization angle $\alpha$ and the phase difference between $p$- and $s$-polarizations $\delta_p - \delta_s$ that maximize the energy transfer from far-field to SPP near-field. While the former is further deconvoluted into the lattice and basis contributions $\alpha_{lattice}$ and $\alpha_{basis}$, the latter is determined by the phase difference between the $p$- and $s$-excited SPPs. Nevertheless, they all are manifested by intrinsic and extrinsic chirality. On one hand, the basis acts like two rectangular slots in which their dimension and separation control the resulting dipole moment, which dominates $\alpha_{basis}$ and $\delta_p - \delta_s$. On the other hand, the relative orientation between the lattice and the incidence governs $\alpha_{lattice}$. In order to maximize $g$, dual slot systems are then proposed. Guided by the CMT model, $g$ as large as 1.82 and 1.55 can be obtained from simulation and experiment. Our results explain the CD effect mediated by single mode SPPs on periodic structures. The formalism can be readily generalized to different periodic structures which support SPPs since the effects of basis and lattice on CD are encoded into $\alpha$ and $\delta_p - \delta_s$. More importantly, it provides a systematic way of rationally designing the system for optimizing CD, which is much better than the conventional trial-and-error approach.

## X. ACKNOWLEDGEMENT

This research was supported by the Chinese University of Hong Kong through Area of Excellence (AoE/P-02/12) and Innovative Technology Funds (ITS/133/19 and UIM/397).

## XI. APPENDIX: DERIVATION OF SCATTERING MATRIX

To solve for the elements in $S(\omega, \vec{k})$, we combine Eq. (5) and (7) and find the following conditions:

$$\tilde{r}_{pp}^{k} = \tilde{r}_{pp}^{-k}, \quad (A1) \qquad \tilde{r}_{ss}^{k} = \tilde{r}_{ss}^{-k}, \quad (A2)$$

$$\tilde{r}_{ps}^{k} = \tilde{r}_{sp}^{-k}, \quad (A3) \qquad \tilde{r}_{sp}^{k} = \tilde{r}_{ps}^{-k}, \quad (A4)$$

$$d_p^k \kappa_p^k = d_p^{-k} \kappa_p^{-k}, \quad (A5) \qquad d_s^k \kappa_s^k = d_s^{-k} \kappa_s^{-k}, \quad (A6)$$

$$d_p^k \kappa_s^k = d_s^{-k} \kappa_p^{-k}, \quad (A7) \qquad \text{and} \qquad d_s^k \kappa_p^k = d_p^{-k} \kappa_s^{-k}. \quad (A8).$$

It is expected that the off-diagonal elements are not the same, i.e. $\tilde{r}_{sp}^k \neq \tilde{r}_{ps}^k$ and $d_p^k \kappa_s^k \neq d_s^k \kappa_p^k$, because $S(\omega, \vec{k})$ itself is not symmetric. In addition, the in- and out-coupling constants are different, i.e. $d_p^k \neq \kappa_p^k$ and $d_s^k \neq \kappa_s^k$. To formulate the in- and out-coupling constants, we consider the time reversal symmetry and assume the incidences are absent so that

$$\frac{d|a^k|^2}{dt} = -\Gamma_{rad}|a^k|^2 = -\left(|s_{-p}^k|^2 + |s_{-s}^k|^2\right) = -\left(|d_p^k|^2 + |d_s^k|^2\right)|a^k|^2 \text{ and } \Gamma_{rad} = |d_p^k|^2 + |d_s^k|^2.$$

We then can write $d_p^k = \sqrt{\Gamma_{rad}} \cos \alpha_k^{out} e^{i\delta_p^{k,out}}$ and $d_s^k = \sqrt{\Gamma_{rad}} \sin \alpha_k^{out} e^{i\delta_s^{k,out}}$, where $\alpha_k^{out}$ and $\delta_{p,s}^{k,out}$ are the out-coupling polarization angle and the $p$- and $s$-out-coupling phase shifts. Similarly, the in-coupling constants are $\kappa_p^k = \sqrt{\Gamma_{rad}} \cos \alpha_k^{in} e^{i\delta_p^{k,in}}$ and $\kappa_s^k = \sqrt{\Gamma_{rad}} \sin \alpha_k^{in} e^{i\delta_s^{k,in}}$. Once all the elements are available, we then compose $S(\omega, \vec{k})$ under the forward direction as in Eq. (8).

Note that combining Eq. (A5) and Eq. (A7), together with Eq. (A5) and Eq. (A8), we can get $\alpha_k^{in} = \alpha_{-k}^{out}$ and $\alpha_k^{out} = \alpha_{-k}^{in}$ respectively. As a result, the backward direction $S(\omega, -\vec{k})$ can be written in a similar fashion following the conditions provided in Eq. (A1) - (A8) together with $\cos \alpha_k^{in} = \cos \alpha_{-k}^{out}$, $\cos \alpha_k^{out} = \cos \alpha_{-k}^{in}$, $\delta_p^{k,in} = \delta_p^{-k,out}$, and $\delta_p^{k,out} = \delta_p^{-k,in}$.

**FIGURES**

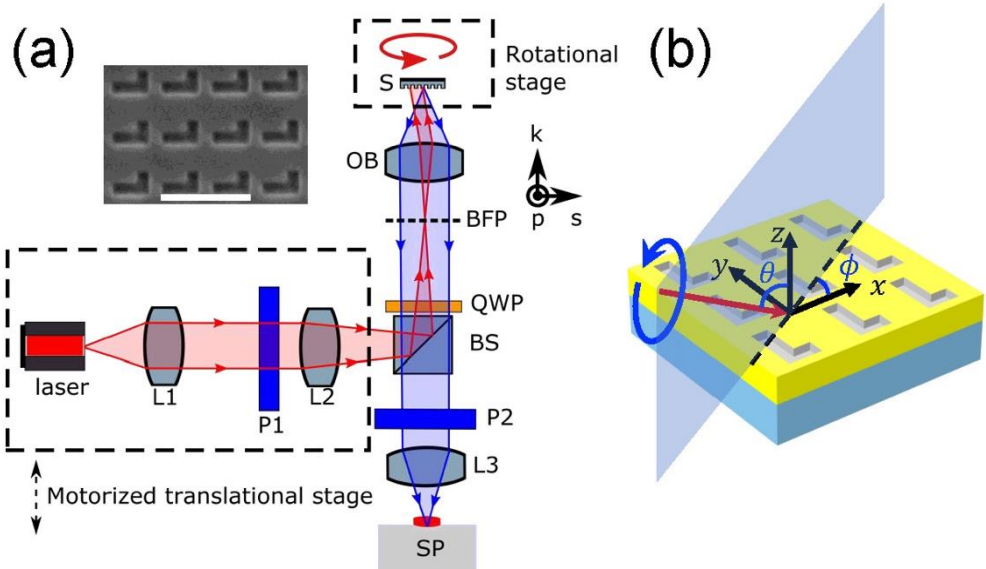

Figure 1. (a) The setup for angle- and polarization-resolved optical microscopy where L: focusing lens, P: polarizer, BS: beamsplitter, QWP: quarter wave plate, BFP: back focal plane, OB: objective lens, S: sample and SP: spectrometer. The red beam indicates the incident beam path and the blue beam defines the scattered beam path from the sample. Inset: the plane-view SEM image of the FIB-fabricated L-shape metallic nanohole array with the scale bar being 1 $\mu m$. (b) The schematic of the sample and the excitation configuration. The blue region is the glass substrate while the yellow region is the gold thin film. The incident polar angle is defined as $\theta$ respect to the surface normal, or the z-direction, along the incident plane and the incident azimuthal angle is defined as $\phi$ between the incident plane and the Γ-X direction, i.e. the x-direction.

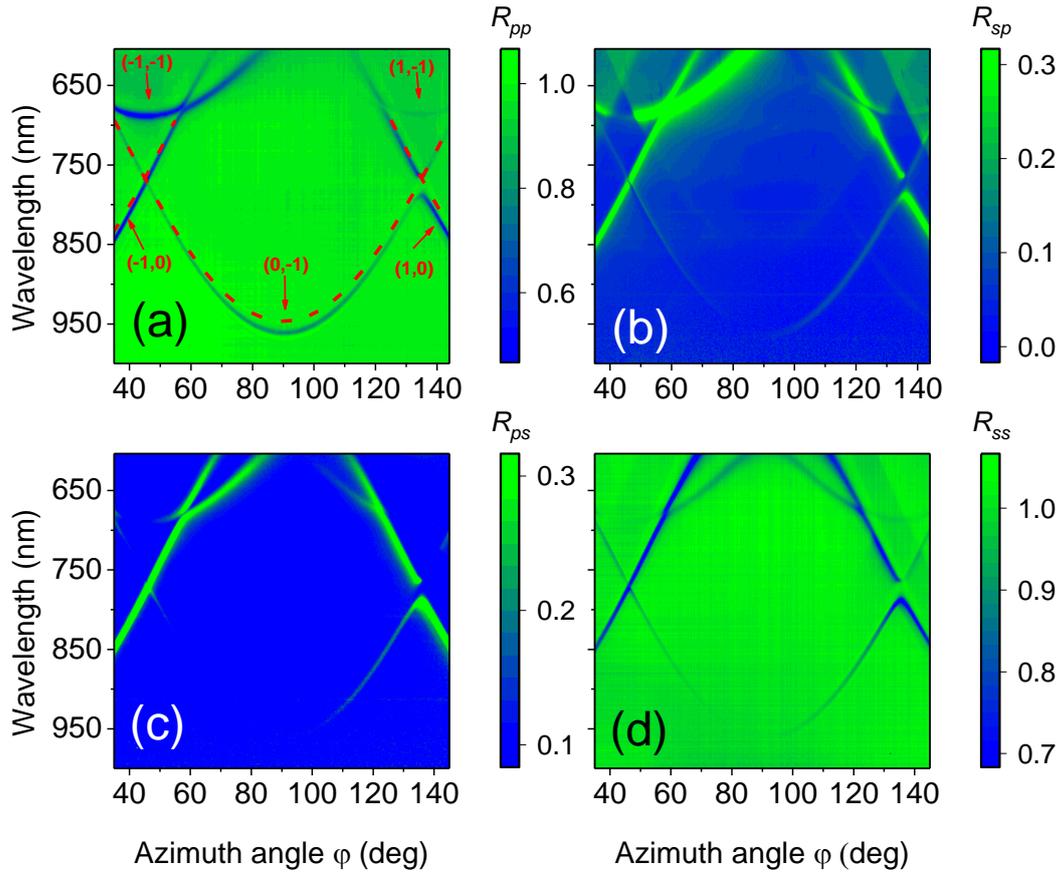

Figure 2. The measured $\phi$-dependent mappings of (a) $R_{pp}$, (b) $R_{sp}$, (c) $R_{ps}$ and (d) $R_{ss}$. The SPP dispersion relations deduced from the analytical phase-matching equation are shown as red dash lines in (a) and are labelled as $(n_x, n_y)$. The mappings are not symmetric with respect to $\phi = 90°$ due to the chiral basis.

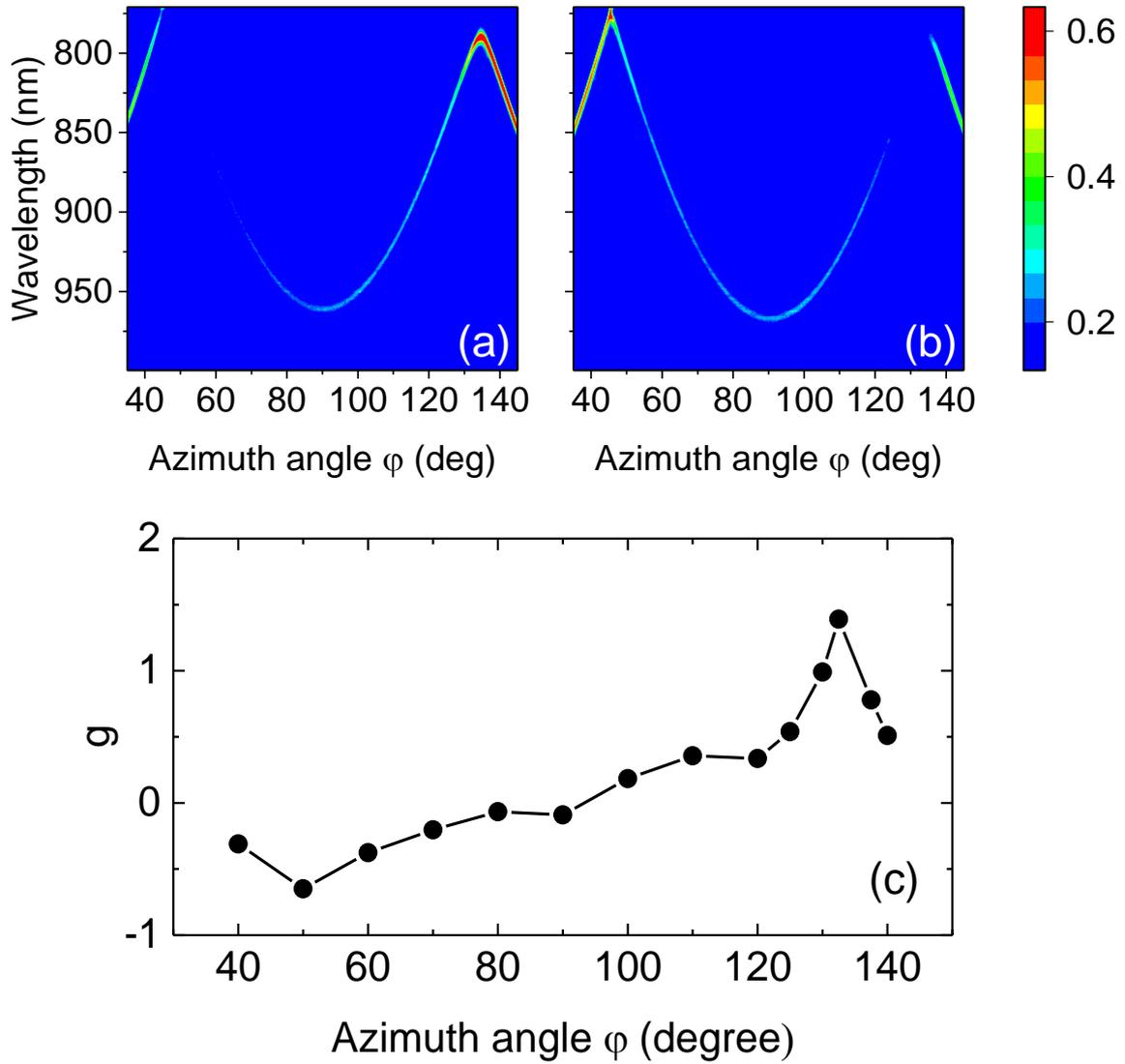

Figure 3. The measured $\phi$-dependent absorption mappings of the (-1,0), (0,-1) and (1,0) SPP modes taken under (a) RCP and (b) LCP excitations. The color scale bars indicate the strength of absorption. (c) The plot of the dissymmetry factor *g* as a function of $\phi$.

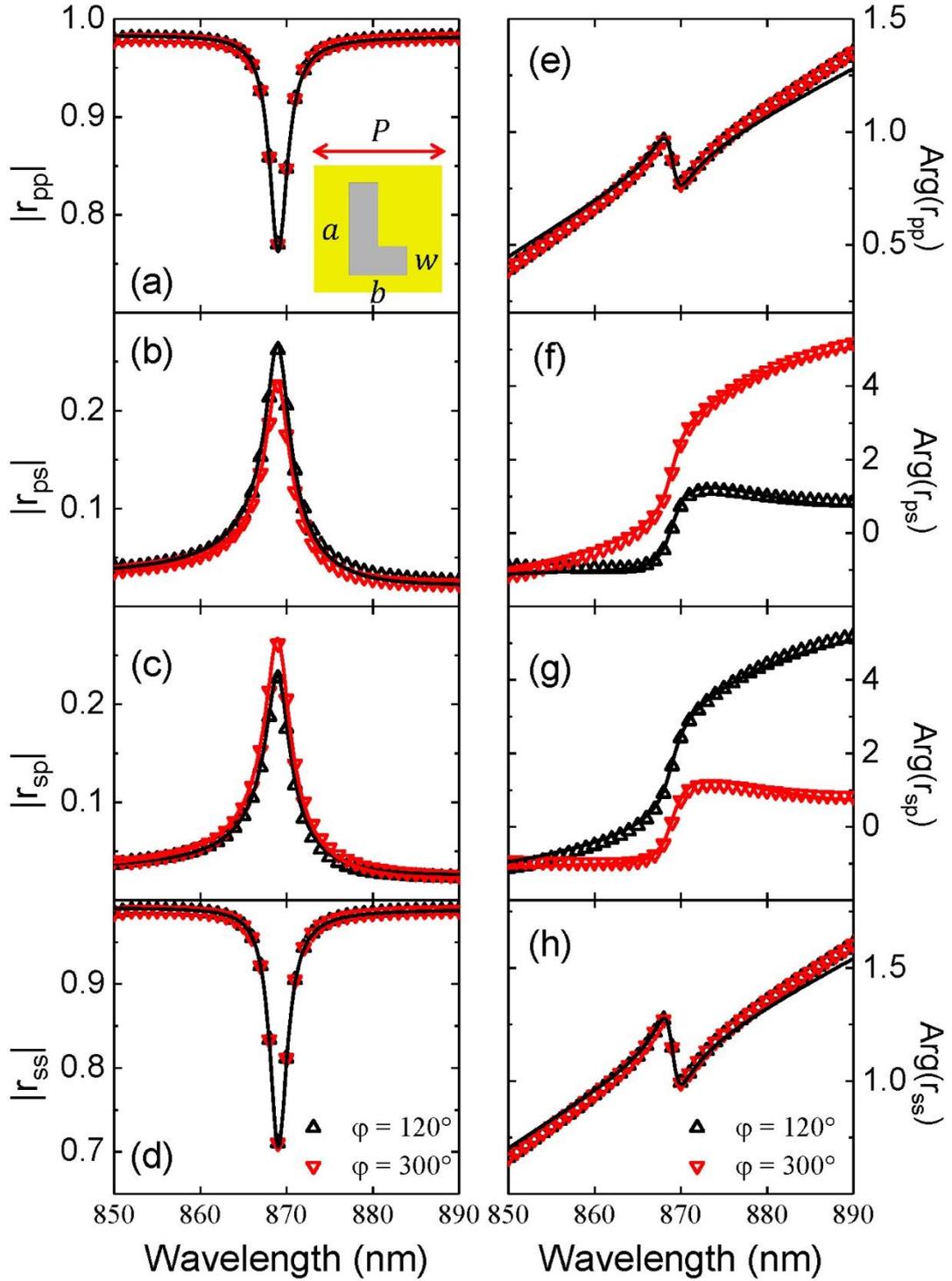

Figure 4. The simulated (a)-(d) amplitudes and (e)-(h) phases of the reflection coefficients $r_{pp}$, $r_{ps}$, $r_{sp}$, and $r_{ss}$ for $\phi = 120°$ (forward incidence) and $300°$ (backward incidence). The solid lines are the best fits by using the CMT deduced $S$. Inset: the schematic of the simulation unit cell with $P = 550$ nm, hole depth = 60 nm, $a$ and $b$ = 400 and 250 nm and $w = 125$ nm.

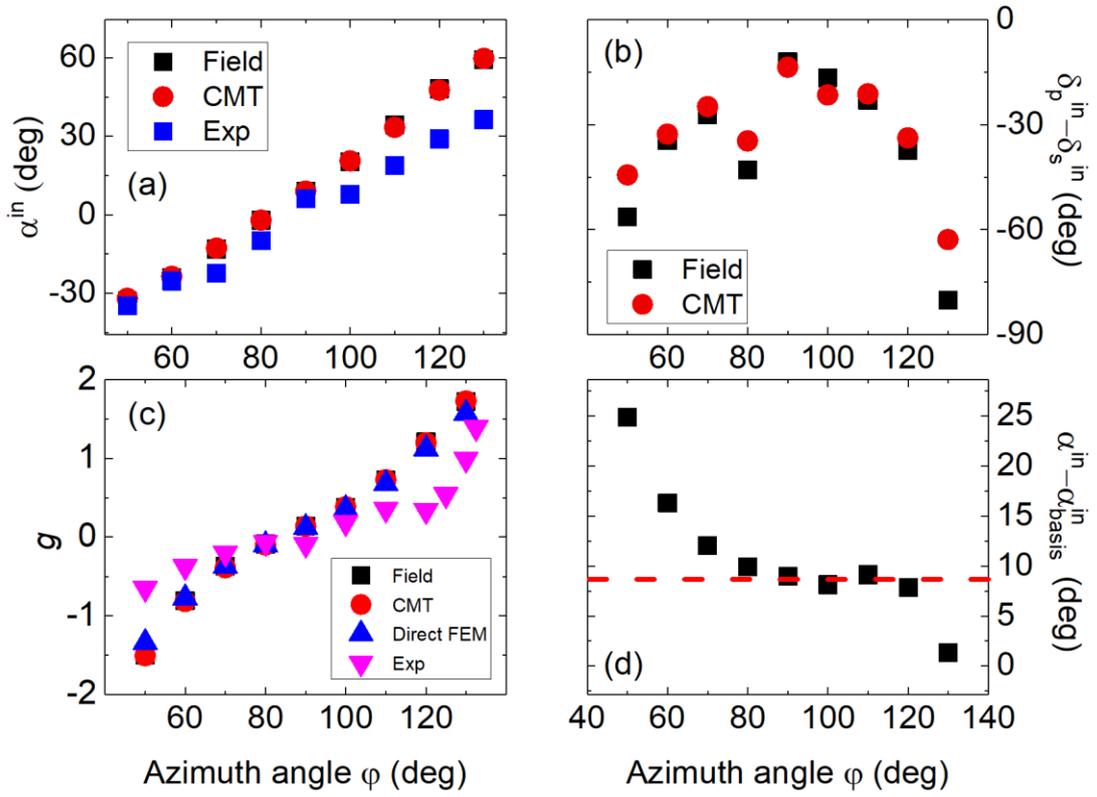

Figure 5. (a) The $\alpha^{in}$-$\phi$ obtained by fitting the (red circles) simulation and the (blue squares) experimental results with the CMT deduced $S$ and by using the field manipulation method (black squares). (b) The $\delta_p^{in}$ - $\delta_s^{in}$ plot obtained by fitting the (red circles) simulation with the CMT deduced $S$ and by using the field manipulation method (black squares). (c) The $g$-$\phi$ calculated by direct FEM (blue triangles) and by using the $\alpha^{in}$ and $\delta_p^{in}$ - $\delta_s^{in}$ given in (a) & (b) from CMT (red circles) and field manipulation (black squares). The experimental results extracted from Fig. 3(c) is also overlaid for comparison. (d) The plot of $\alpha^{in} - \alpha^{in}_{basis}$ as function of $\phi$. It is almost constant with $\phi$ when away from the cross-points.

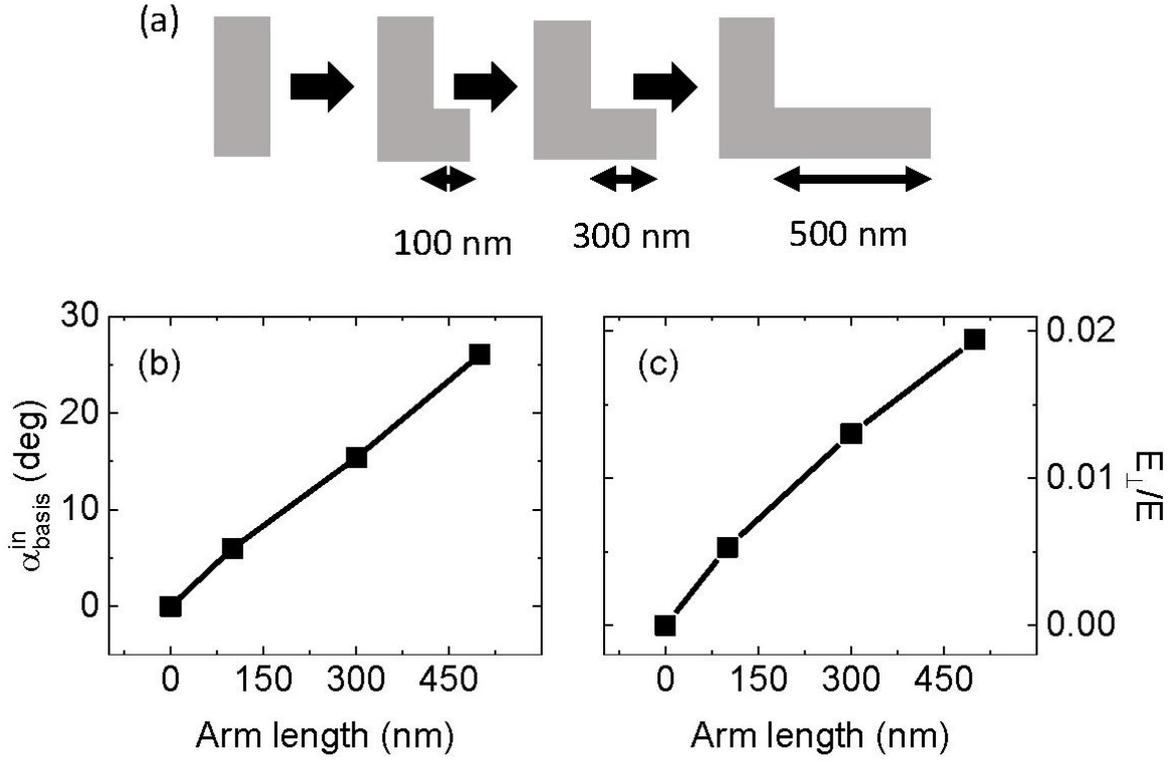

Figure 6. (a) The evolution of the basis with arm length along the Γ-X direction. The arm length $b$ is varied from 0 to 500 nm. (b) The variation of $\alpha^{in}_{basis}$ with $b$ provided $\alpha^{in}_{lattice}$ is almost 0° all the time. (c) The plot of the ratio of the field amplitude perpendicular to the SPPs propagation direction and the total field amplitude as a function of $b$, showing a consistent behavior with $\alpha^{in}_{basis}$.

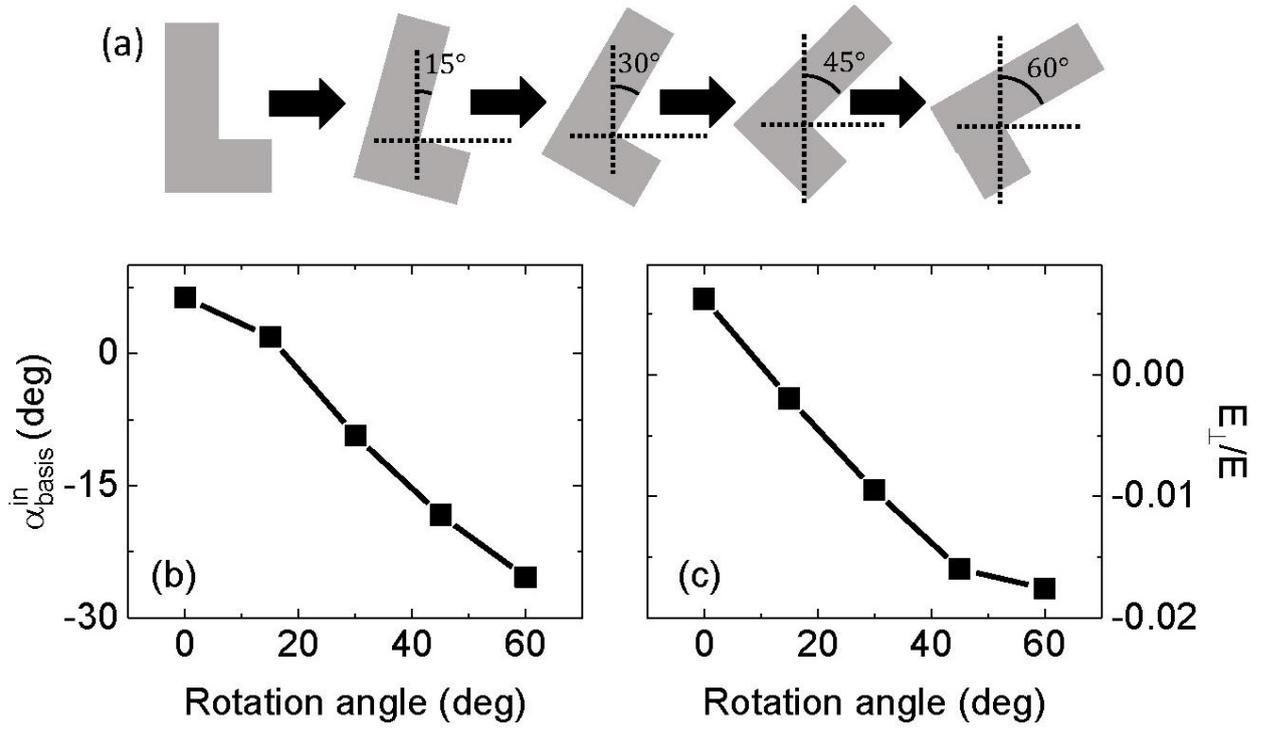

Figure 7. (a) The evolution of the basis with orientation along the Γ-X direction. The rotation angle is varied from 0 to 60° with respect to the vertical axis. (b) The variation of $\alpha^{in}_{basis}$ with rotation angle provided $\alpha^{in}_{lattice}$ is 0° all the time. (c) The plot of the ratio of the field amplitude perpendicular to the SPPs propagation direction and the total field amplitude as a function of the rotation angle, showing a consistent trend.

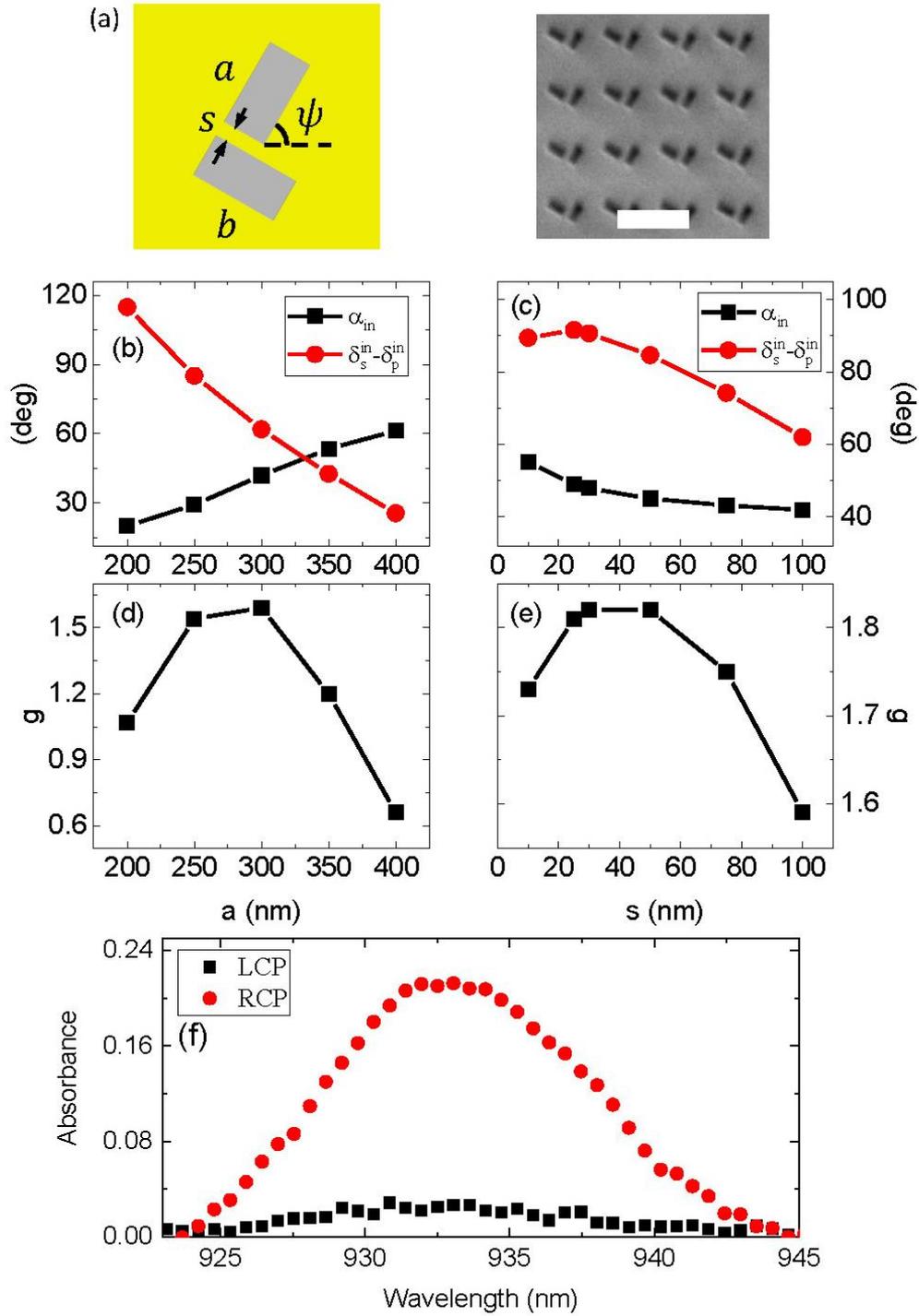

Figure 8. (a) The unit cell of the proposed dual slot structure and the plane-view SEM image of the dual slot periodic structure fabricated by FIB with the scale bar = 1 $\mu m$. The plots of simulated (b) $\alpha^{in}$ and $\delta_p^{in}$ - $\delta_s^{in}$ and (d) $g$ as a function of $a$. The plots of simulated (c) $\alpha^{in}$ and $\delta_p^{in}$ - $\delta_s^{in}$ and (e) $g$ as a function of $s$. The highest $g$ obtained from two series are 1.52 and 1.82, respectively. (f) The absorption spectra of the (0,-1) SPP mode measured under LCP and RCP excitations, showing the $g$ reaches 1.55.

# Supplementary Information

# Generalization of the circular dichroism from metallic arrays that support Bloch-like surface plasmon polaritons

X. Guo, C. Liu, and H.C. Ong

Department of Physics, The Chinese University of Hong Kong, Shatin, Hong Kong, People's Republic of China

I. The simulated and best fitted reflection spectra of the Au array

The complex reflection coefficient $r$, amplitude and phase, spectra of the (0,-1) SPPs from the L-shape Au array simulated by COMSOL at incident polar angle $\theta = 45°$ and different azimuthal angles $\phi$ from 50° to 130° (symbols) and then best fitted by the analytical scattering matrix $S$ deduced by the temporal coupled mode theory (solid lines). The simulated unit cell has $P = 550$ nm, hole depth = 60 nm, $a$ and $b$ = 400 and 250 nm and $w = 125$ nm.

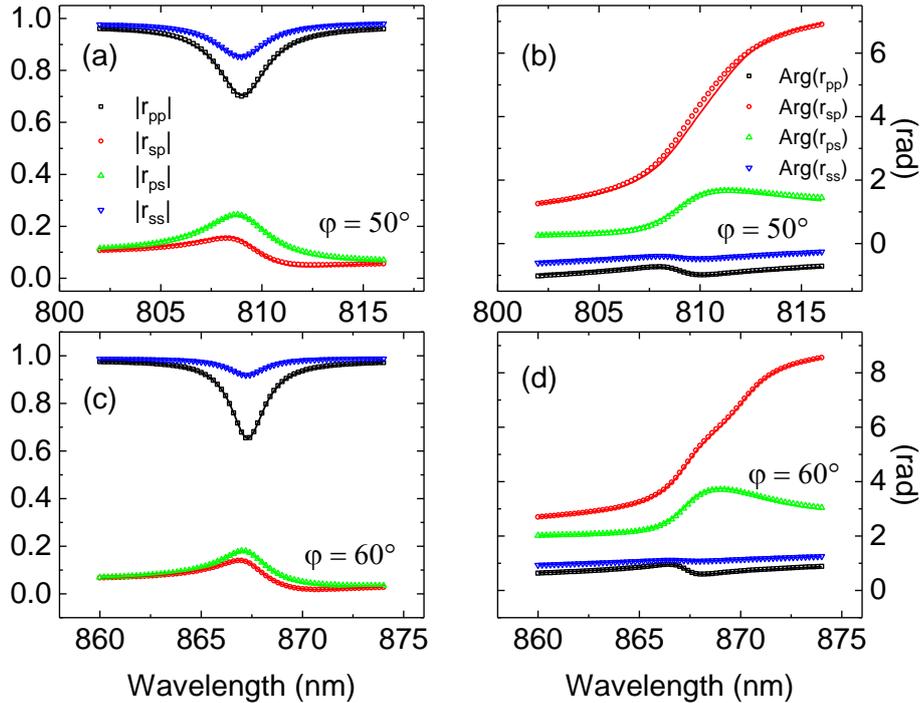

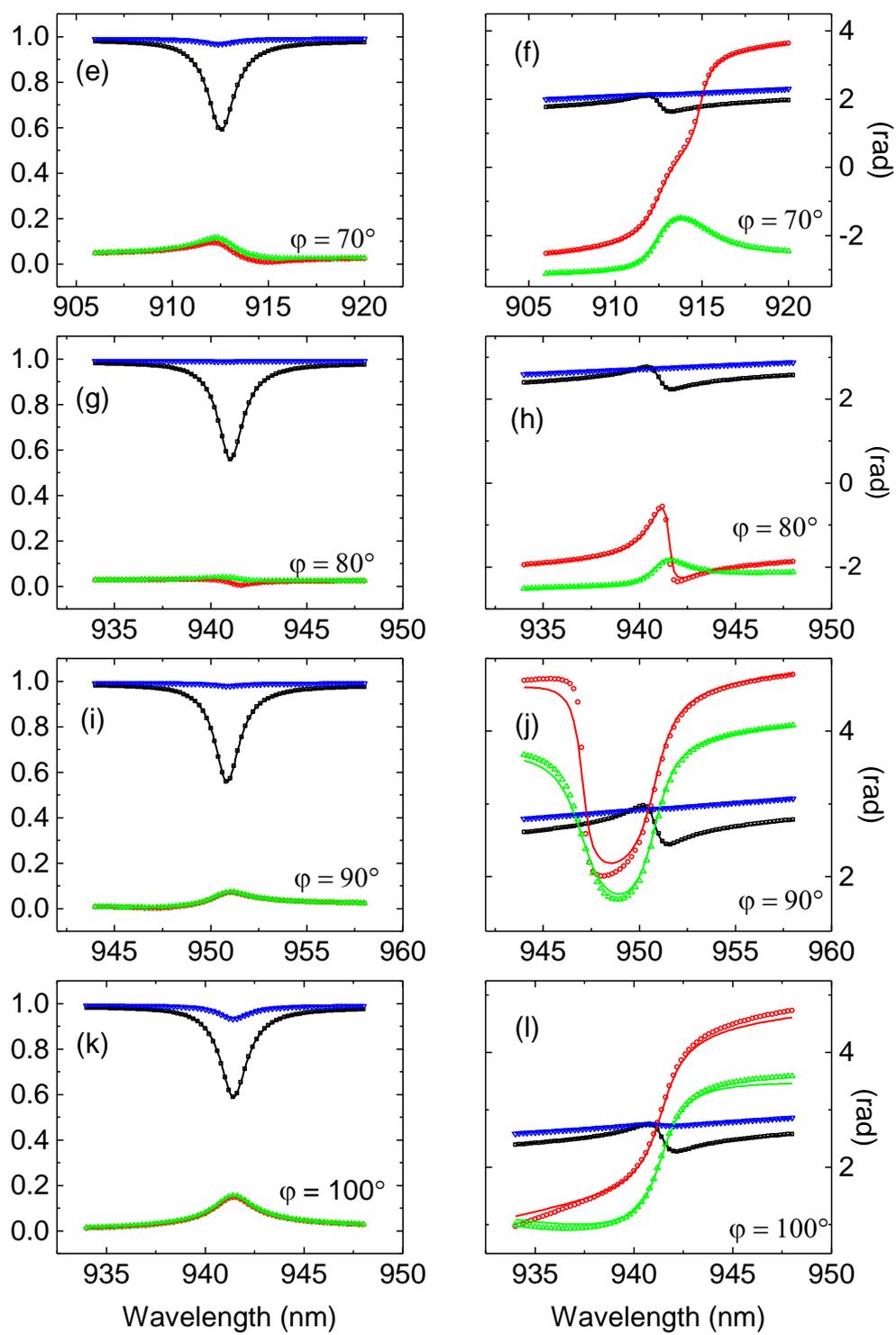

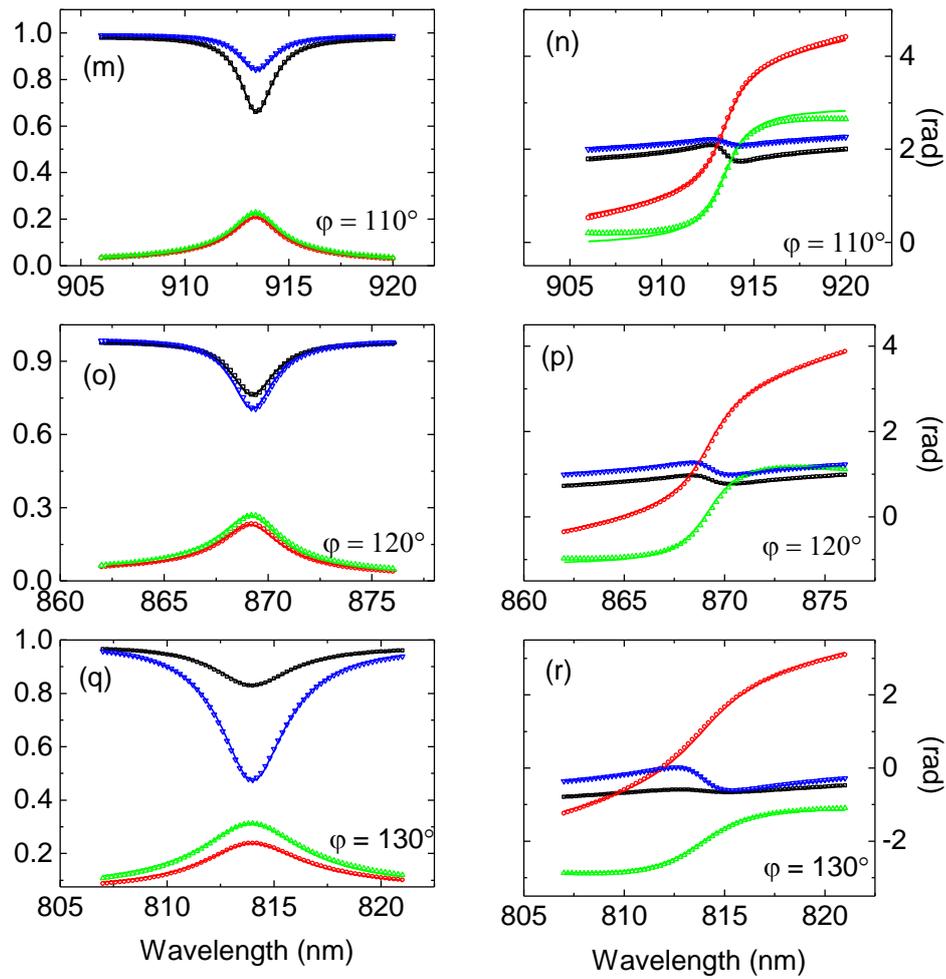

Figure 1S. The simulated amplitudes (left column) and phases (right column) of the reflection coefficients $r_{pp}$, $r_{ps}$, $r_{sp}$, and $r_{ss}$. Solid lines indicate the best fits.

II. The simulated ($\alpha^{in} - \alpha^{in}_{lattice}$) - $\phi$ plot from another L-shaped nanohole array.

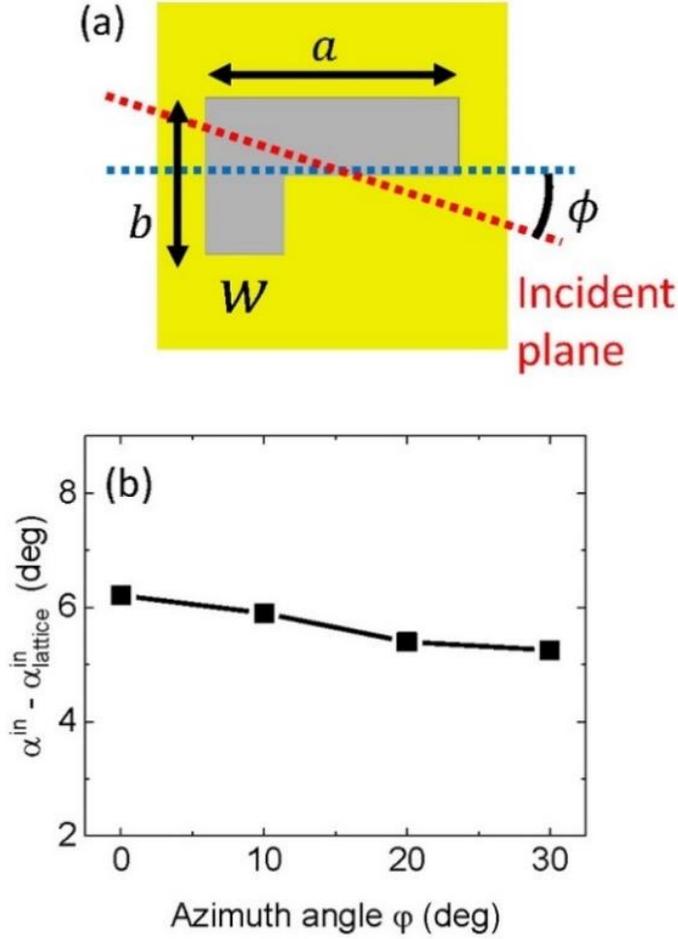

Figure 2S. (a) The schematic of the L-shaped nanohole array unit cell and the excitation configuration. The array has $P = 800$ nm, $H = 60$ nm, $a$ and $b = 400$ and 250 nm as well as $w = 125$ nm taken at $\theta = 5°$. (b) The variation of $\alpha^{in} - \alpha^{in}_{lattice}$ as a function of $\phi$ simulated by COMSOL.

It can be observed from Fig. 2S(b) that $\alpha^{in} - \alpha^{in}_{lattice}$ is almost a constant between 5° to 6°. At $\theta = 5°$, following the phase-matching equation, the SPPs propagate primarily along the Γ-X direction. As $\alpha^{in}$ indicates the overlapping of the incident polarization and the electric field of SPPs, which aligns with the propgation direction, the constant discrepancy between $\alpha^{in}$ and $\alpha^{in}_{lattice}$ suggests there exists an additional factor that effectively perturbs the field of SPPs. Knowing both the basis and the propagation direction of SPPs remian unchanged as a fucntion of azimuthal angle, we thus propose the additional factor arises from the basis that produces a dipole moment which is orthogonal to the SPP propagation and $\alpha^{in} = \alpha^{in}_{lattice} + \alpha^{in}_{basis}$.

III.     The electric near field patterns of the dual slot excited by *p*- and *s*-polarized lights.

The electric field patterns of the dual slot excited by *p*- and *s*-polarized light are simulated by COMSOL under off-resonance to avoid the excitation of SPPs. The off-resonance excitations show two slots respond differently to *p*- and *s*-polarizations. While the lower slot responds only to the *p*-light, the upper slot is excited mainly by the *s*-light. As a result, when under an elliptically polarized light, two slots are decoupled, and each excites its own SPPs. The two SPPs then interfere with their amplitudes and relative phase difference controlled independently by the input polarization.

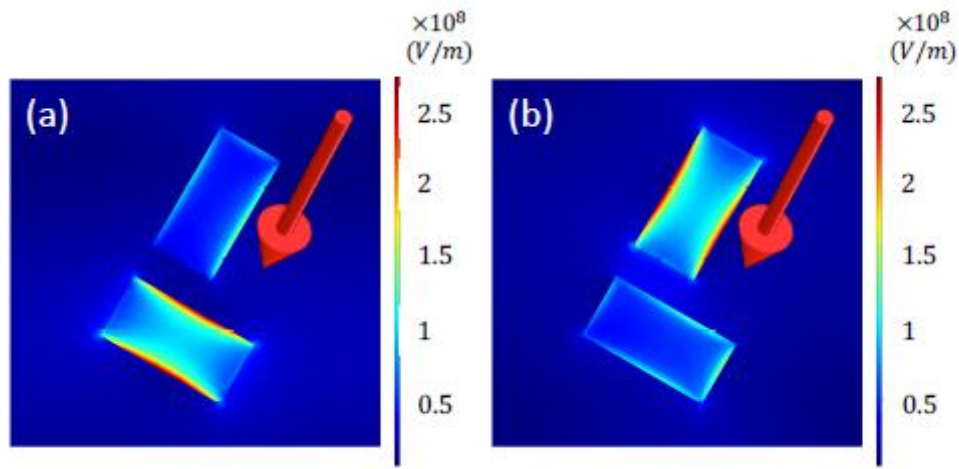

Figure 3S. (a) Under *p*-polarized excitation, the off-resonance electric field localizes at the lower slit, while (b) under *s*-polarized excitation, the off-resonance electric field localizes at the upper slit. Red arrows indicate the incident light directions. $\theta = 10°$, $\phi = -120°$ and $\lambda = 1000$ nm.